
\magnification=1200
\font\abs=cmss10
\hsize=6truein
\vsize=9.5truein
\null \vskip 1truecm
{\nopagenumbers

\rightline{ CRN-94-17}
\rightline{ March 1994}
\vskip 0.5truecm
\centerline{{\bf  CLIFFORD ALGEBRAS OF POLYNOMIALS}}
\vskip 1truecm
\centerline{{\bf  GENERALIZED GRASSMANN ALGEBRAS}}
\vskip 1truecm
\centerline{{\bf  and}}
\vskip 1truecm
\centerline{{\bf  q-DEFORMED HEISENBERG ALGEBRAS}}
\vskip 3truecm

\centerline{M.~Rausch de Traubenberg }
\bigskip
\bigskip
\centerline{Physique Th\'eorique}
\centerline{Centre de Recherches Nucl\'eaires et Universit\'e Louis
Pasteur}
\centerline{B.P. 20 F-67037 Strasbourg Cedex}
\vskip 1truecm
\hfuzz=1pt
\hfuzz=1pt
\vskip 2.5truecm
\noindent
\underbar{ABSTRACT}
\bigskip
\noindent
{\abs
Classification of finite dimensional representations of the q-deformed
Heisenberg algebra $H_q(3)$ is made by the help of Clifford algebra of
polynomials
and generalized Grassmann algebra.
Special attention is paid when $q$ is a primitive $n$ root
of unity. As a further application we obtain  finite dimensional
representations of $ sl(2)_q$ using its  embedding into $H_q(3)$.
}

\vfill\eject}
\bigskip
\noindent
\bigskip
Symmetries have always played a central role in physics, however under
special assumptions it is interesting to make some deformations of
the basic relations of the group of symmetry.
These deformations  should be
understood as perturbations of the  symmetry. There are two ways of making
such deformations: on the group of symmetry, either on its associated
universal enveloping algebra [1]; or directly on the algebra itself [2]
(see also [3]). The first type of deformations, also called quantum groups,
has a Hopf algebra structure [1], whereas for the second type (quantum
algebras)
it is not necessarily the case .
The consideration of such algebras is
motivated, for example,  by the resolution of the Yang-Baxter equation,
which appears  in particle or statistical physics [4] amongst others.
The quantum algebras, deformed by an $n$ primitive root of
unity $\omega$ should have a $Z_n$ graded structure [2].
It is known that superalgebras [5]
( $Z_2$-graded algebras) are subtended by Clifford algebras or Grassmann
algebras or, in other words, by quadratic forms. Similarily the $\omega$-
deformed algebras  may be understood within the framework of Clifford
algebras of polynomials [6] $C_f$, or $n$-exterior algebra [7]
$\Lambda_n$. These algebras were introduced to allow a linearization of the
$n$-degree polynomial $f$ for the former, and to take the $n$-root of
the null polynomial for the latter. But, in contrast to  the usual
Clifford or Grassmann cases, these algebras are not finite [6,7,8].
However, it has been shown that a finite ( but not faithful) representation
can be obtained
for cubic polynomials [9] or even general ones [10]. The main
property of the $C_f$ algebra is that it can be connected naturally to the
n-exterior algebra [11]. This algebra has, as a finite
(not faithful) representation, the generalized Grassmann algebra [10,12],
which is the fundamental tool to obtain  finite dimensional representations
of $H_\omega(3)$, the $\omega$-deformed Heisenberg algebra [13].
 Let us mention that an n-dimensional representation
has already been obtained in [14]. The content of this paper is
organised as follows. In sect.1 we recall the basic and useful properties of
Clifford algebras of polynomials. Construction of
finite dimensional representations of $H_q(3)$ is carried out
 in sect.2, and that of
 $sl(2)_q$ in sect.3.
 \vskip 1truecm
\noindent
1 -- \underbar{Clifford algebras of polynomials}
\vskip 1truecm
\noindent
Let  $f$ be a homogeneous polynomial of degree n with p variables.
Its associated $C_f$ is generated by the p elements $g_i$ ($i=1 \cdots p$)
allowing the linearization of $f$ [6]
$$\eqalign{
f(x)&=
\sum \limits_{\{i\}=1}^p
x_{i_1} \cdots x_{i_n}\;g_{i_1 \cdots i_n} \cr
&=(x_1 g_1 + \cdots + x_p g_p)^n },\eqno(1.1.a)$$
where $g_{i_1 \cdots i_n}$ is the symmetric tensor associated with $f$.
By developing the $n$-th power, the result is that the generators $g_i$
of $C_f$ are
submitted to the constraints
$$ {1\over n!} \sum\limits_{\sigma \in \Sigma_n}
 g_{i_{\sigma(1)}} \cdots g_{i_{\sigma(n)}}=
g_{i_1 \cdots i_n}\eqno(1.1.b)$$
where $\Sigma_n$ is the group of permutation of n elements. In spite of the
fact that all the $C_f$ algebras are not equivalent, we can naturally
 associate  the n-exterior algebra $\Lambda_n^{(p)}$ to $C_f$ [11].
It is defined by taking $f(x) = 0 $ in (1.1.a), and hence  by the following
constraints
$$ \sum\limits_{\sigma \in \Sigma_n}
g_{i_{\sigma(1)}} \cdots g_{i_{\sigma(n)}}= 0\eqno(1.2)$$
Like $C_f, \Lambda_n^{(p)}$ is  of infinite dimension, but a finite dimensional
representation can be obtained through the homomorphism of
$ \Lambda_n^{(p)}$ on   $G_n^{(p)}$ which maps
$g_i$ in    $\theta_i$ in such a way that
$0 \neq g_ig_j - \omega g_i g_j ( i<j )$ is sent to  zero [10,12].
The $n^p$-dimensional algebra so obtained is generated by the p
canonical generators fulfilling
$$\eqalign{\theta_i \theta_j &=\omega \theta_j \theta_j\;\;\;(i < j )\cr
\theta_i^n&=0\cr}\eqno(1.3) $$
It can be proved that the $\theta$ 's satisfy (1.2) as it should  for
a representation of $\Lambda_n^{(p)}$, but  (1.2) does not lead
to (1.3). It has also been observed that the minimal faithful representation
of $G_n^{(p)}$ is obtained by $n^p \times n^p$matrices [10,12]. This algebra,
also called paragrassmann\footnote{*}
{Strictly speaking, for a paragrassmann algebra, there are  cubic
constraints instead of (1.3) [15]}
 appears quite naturally in the frame of
parastatistics [15], parasupersymmetric quantum mechanics, parasuperalgebras
 [16]
and even
in $2D$ conformal field theory [17]. The correspondance between
$C_f$ and $\Lambda$ can also be obtained by an explicit calculation as
we will see. The fundamental property which leads to an explicit
matrix representation of $C_f$ is based on the following property: to
linearize any polynomial, only two basic polynomials have to be considered
- the sum $ S(x)= x_1^n + \cdots + x_p^n$ and the product
$P(x)=x_1 \cdots x_n$ - which are linearized by matrices that turn out
to be a representation of the generalized Clifford algebra [10]
$C_n^{(p)}$. This $n^p$-dimensional algebra is defined by p
canonical generators $\gamma_i$ satisfying [18]

$$\eqalign{\gamma_i \gamma_j &=\omega \gamma_j \gamma_j\;\;\;(i < j )\cr
\gamma_i^n&=1\cr}\eqno(1.4) $$

\noindent
and has a large range of applications in physics (see [10,12]
for references), and leads to a finite dimensional quantum mechanics
[19] ( see also refs.[8,9] of [12]). This last property has to be compared
with the next sect. where we study the $\omega$- deformed
Heisenberg algebra. The minimal faithful representation of the
$\gamma_i$ is obtained with $ n^{k} \times n^{k}$ matrices, $k=E(p/2)$
$$\eqalign{
\gamma_{2i-1}&=\sigma_3^{\otimes(i-1)}\otimes\sigma_1\otimes1^{\otimes(k-i)}\cr
 \gamma_{2i}
&=\sigma_3^{\otimes(i-1)}\otimes\sigma_2\otimes1^{\otimes(k-i)}\cr
 \gamma_{2p+1}&=\sigma_3^{\otimes(k)}}\eqno(1.5)$$
where $i = 1 \cdots p$, 1 is the $n \times n$ identity matrix and

$$\sigma_1=\pmatrix{
0&1&&&\cr &\ddots&\ddots&&\cr &&\ddots&\ddots&\cr &&&\ddots&1\cr
1&\cdots&\cdots&\cdots&0\cr}
\qquad\sigma_3=\pmatrix{
\omega &&0&\cr &\omega^2&&\cr &&\ddots&\cr
 &0&&1\cr}\eqno(1.6.a)$$

 $$
\sigma_2=\cases{\sigma_3\sigma_1&($n$ odd)\cr
\sqrt\omega \sigma_3\sigma_1&($n$ even)\quad
with~$\omega=\exp\left({2i\pi\over n}\right)$.\cr}\eqno(1.6.b)$$

\noindent
Similar formulas can be found in [10,12,18].
{}From the generalized Clifford algebra
 $C_n^{(2p)}$, one can build n explicit $G_{n,l}^{(p)}$
generalized Grassmann  algebras  $(l=0 \cdots n-1)$ [10,12]
$$ \theta_i^{(l)} = \gamma_{2i-1} + \omega^{l+{1\over 2} }\gamma_{2i}
\eqno(1.7)$$
and the reciprocal formulas are
$$\eqalign{
\gamma_{2i-1}=& \sum\limits_{l=0}^{n-1} \theta_i^{(l)} \cr
\gamma_{2i}  =& \sum\limits_{l=0}^{n-1} \omega^{-l-{1\over 2} } \theta_i^{(l)}}
\eqno(1.8)$$
Due to the property stating that finite dimensional representations
of $C_f$ are subtended by $C_n^{(p)}$, and due to (1.7), we see that
any polynomial can be naturally associated to $G_n^{(p)}$ by the following
sequence $f \rightarrow C_f \rightarrow \Lambda_n \rightarrow G_n$.
The purpose of the
next sect. is to connect  $G_n$ with the $\omega$-deformed Heisenberg algebra
$H_{\omega}(3)$.
\vskip 0.3truecm
\noindent
It is worth noting that generalized Clifford or Grassmann algebras constitute
a representation of the quantum hyperplain $ \Re_q^p$, with an appropriate
$R$ matrix [20]
$$R^{i_1i_2}_{\;\;\;\;\;\;j_1j_2} = \delta^{i_2}_{\;\;\;j_1}
\delta^{i_1}_{\;\;\;j_2}(1 + (\omega - 1) \delta^{i_1i_2})
  +(\omega - \omega^{-1})\delta^{i_1}_{\;\;\;j_1}
   \delta^{i_2}_{\;\;\;j_2} \Theta(i_2-i_1) \eqno(1.9) $$
where $\Theta(i_2-i_1)$ is equal to  $1$ if $i_2>i_1$, otherwise $0$.

\noindent
In term of the $R$ matrix, the equations (1.3.a) and (1.4.a) become

$$ (\omega^{-1}R_{12} - 1_{12})x_1x_2 = 0 \eqno(1.10)$$
where $x$ belongs to $\Re_q^p$, $R$ is understood as a matrix which represents
an application of $\Re_q^p \otimes \Re_q^p$ in itself, and $1_{12}$ is
the identity
matrix. A direct calculation  shows that $R$ is a solution of the
Yang-Baxter equation
$$ R_{12} R_{23} R_{12} = R_{23} R_{12} R_{23} \eqno(1.11)$$
which indicates how applications of $\Re_q^p \otimes \Re_q^p \otimes \Re_q^p$
can be treated.

\vskip 0.3truecm
\noindent
It is possible to construct an analysis compatible with the $Z_n$
graded structure of $G_n$, the results obtained in [14] for p=1 should be
recalled briefly.
The $Z_n$ graded Leibnitz rule
$$\partial(ab)=\partial(a)b+\omega^{gr(a)}a\partial(b) \eqno(1.12)$$
where $gr(\theta^p)=p, gr(\partial^p)=-p$, leads to
$$\partial(\theta^p) = \{p\} \theta^{p-1} \eqno(1.13)$$
with $\{p\} = {(1 -\omega^p)\over(1-\omega)}$. Using (1.5) and (1.7),
 introducing
$\partial = F \theta F^{-1}$ ( finite Fourier transformation)
 and utilizing (1.8), we
get a matrix representation of $H_{\omega}(3)$
$$\partial \theta - \omega \theta \partial = 1 \eqno(1.14)$$
\vskip 1truecm
\noindent
2 -- \underbar{q- deformed Heisenberg algebras}
\vskip 1truecm
\noindent
The q-deformed Heisenberg algebra is generated by two generators P and
Q satisfying [13]
$$PQ-q QP=1 \eqno(2.1)$$
For $q=1$ , with $Q=x$ and $P=\partial_x$, the usual number and derivative,
 the Heisenberg algebra is obtained, which  has only infinite dimensional
representations in accordance with Bose-Einstein statistics. For $q=-1$
with $Q=\theta$ and $P=\partial_{\theta}$, two Grassmann variables, we
obtain two dimensional representation in accordance with the Pauli exclusion
principle. It is  known however that we can construct statistics which describe
neither fermions nor bosons but parafermions or parabosons [15];
among those two statistics only the former has finite dimensional
representations. This feature has been exploited to build a parasupersymmetric
extension of quantum mechanics [16]. So it seems natural to study finite
dimensional representations of $H_{\omega}(3)$ induced by $G_n^{(1)}$; an
idea which is stressed by the results of H. Weyl [19] connected to
eqs.(1.4-7).
Our starting point is then the resolution of
$$\eqalign{
            Q^n &=0 \cr
            P^n &=0 \cr
    PQ-\omega QP &=1}
            \eqno(2.2)$$
where we want  $Q=\theta$  to belong to $G_n^{(1)}$ and
$P=\partial$, the canonical variable associated with
$\theta$. $n$ is the minimum power of $\theta$ and
$\partial$ which is equal to zero. It is obvious that the minimal size of
$\theta$ is a $n \times n$ matrix (see [10,12] for more details).
\vskip 0.5 truecm
\noindent
a--\underbar{n-dimensional representations.}
\vskip 0.5 truecm
\noindent
We are looking for $n \times n$ matrices satisfying (2.2). If $\theta^n=0$,
it is obvious that all its eigenvalues are zero. Using the Jordan
decomposition of matrices with degenerated eigenvalues, the
only non-zero matrix elements of $\theta$ are those which are under the
principal diagonal. Arguing that $\theta^p \neq  0 ( p < n )$, all
those elements are one. Thus  $\theta$ is always equivalent to
$$\theta = \pmatrix{
          0&0&&0\cr
          1&0&&\cr
          &\ddots&\ddots \cr
           &&1&0}
           \eqno(2.3)$$

\noindent
Having defined $\theta$, we are looking for a derivative satisfying
(2.1) and which is conjugated to $\theta: \partial=F \theta F^{-1}$ where $F$
is
the finite Fourier transformation which must be specified. To obtain
$F$, we proceed in two steps; the first one allows to go from
$\theta$ to $\theta^+$, or in other words to $\sigma_1^+$ to
$\sigma_1$, the second one allows a reproduction of the derivative properties
(1.13). The first step is nothing more than the product of two
Sylvester matrices which transforms $\sigma_1^+$ to $\sigma_3$ [21]
and then $\sigma_3$ to $\sigma_1$. Finally we obtain
$$ F_{ij} = {1\over{\{i-1\}!}} \delta_{i+j-n-1,0} \eqno(2.4)$$
where $\{a\}!=\{a\}\{a-1\} \cdots \{1\}$
and
$$ \partial_{ij} =\{i\} \delta_{i+1j} \eqno(2.5)$$
It is now easy to verify that $\theta$ and $\partial$ generate
$H_{\omega}(3)$ as it should be. At this stage, we have recaptured the matrix
representation obtained in [14], but we know now that it is the only
n-dimensional representation. All the other representations $\theta',
\partial',$ are related to $\theta$ and $\partial$ by an invertible
transformation $P$:$\theta'=P \theta P^{-1},\partial' = P \partial P^{-1}$
and $F'=P F$.
\vskip 0.5truecm
\noindent
b-- \underbar{$k > n $-dimensional representations.}
\vskip 0.5truecm
\noindent
Now we look for $ k \times k$ matrices which satisfy (2.2).
 Using the results
obtained in the previous sub-section as well as  the fact that the minimal
representation of $\theta$ is n-dimensional, we see that we have solutions
of (2.2) iff $k$ is a multiple of $n$ ($k=ln$)

$$\theta =1\otimes \pmatrix{
          0&0&&0\cr
          1&0&&\cr
          &\ddots&\ddots \cr
           &&1&0}
           \eqno(2.6.a)$$

$$\partial =1\otimes \pmatrix{
          0&\{1\}&&\ldots&0\cr
          0&0&\{2\}&\ldots& 0\cr
          &&\ddots&\ddots&\cr
          &&&0&\{n-1\} \cr
           0&&&&0}
           \eqno(2.6.b)$$

\noindent
$1$ is the unit $ l \times l $ matrix. Thus any higher dimensional
representations
are not irreducible and consist of $l$ copies of the smallest representation.

\vskip 1truecm
\noindent
Until  now we have considered the $\omega$-deformed Heisenberg algebra
when $\omega$ is an n-th primitive root of unity.
If it is assumed that
 $ n=n_1n_2$,
using the results obtained previously,  the n-dimensional
representation for
$H_{\omega^{n_{1}}}(3)$ can be built.
 All  the results can be used, but now $\partial$
is a matrix constructed with $\{p\} ={(1-\omega^{pn_1})\over(1-\omega)}$ so
 $$\partial =\pmatrix{
           0&\{1\}&&\ldots&0\cr
          0&0&\{2\}&\ldots& 0\cr
          &&\ddots&\ddots&\cr
          &&&0&\{n_2-1\} \cr
           0&&&&0}
           \otimes 1
           \eqno(2.7)$$

\noindent
where $1$ is the $n_1 \times n_1$ unit matrix.
Thus we have

$$\eqalign{
\partial^{n_2}&=0 \cr
\theta^{n_1n_2}&=0 \cr
\partial \theta - \omega^{n_1} \theta \partial&=0} \eqno(2.8) $$
\noindent
instead of (2.2). In addition, $\partial$ and $\theta$ are not
conjugated in this
case because  $F$ is singular ( see (2.4) ). We  see that some
kind of singular representations are spanned by non-prime numbers.
 Some differences between
prime and non-prime numbers have been noticed in [14].
\noindent
\vskip 0.5truecm
c-- \underbar {Other representations of $H_q(3)$}\footnote{*}
{This sub-section was accomplished  with the help of M. Rosso}
\vskip 0.5truecm
Until now we have simply focused on representations of the
q-deformed Heisenberg algebra induced by  generalized Grassmann
algebra. Now let us consider other representations: it is easy to check
that (2.1) leads to

$$\eqalign{
  PQ^a&=\omega^aQ^aP + \{a\}Q^{a-1}\cr
  P^aQ&=\omega^aP^aQ + \{a\}P^{a-1}}\;\;\;,\eqno(2.9)$$
so when $a=n$ we have
 $$\eqalign{
  PQ^n&=Q^nP \cr
  P^nQ&=QP^n }\;\;\;,\eqno(2.10)$$
\noindent
and considering Shur's lemma, $P^n, Q^n$,  both are proportionnal to
the identity matrix for
irreducible representation. One can then find a basis where $P$ is diagonal
(Cayley-Hamilton theorem)
$$P=\pmatrix{
p_1.I_{r_1}&&&\cr
& p_2.I_{r_2}&&\cr
&&\ddots&\cr
&&&p_k.I_{r_k} }.\eqno(2.11)$$
\noindent where $p_1,\cdots ,p_k$, are the eigenvalues of $P$, which
are degenerated $r_1, \cdots ,r_k$ times. We have assumed that the
dimension of the representation is $r_1 + \cdots + r_k$.
 All the $p_i^{\;\;\;n}$ are equal, and in the special case of
$p_i=0$, we can recapture  the representation of the previous sub-section.
 Inserting this $P$ into
(2.1) resulting in $Q$ ($p_i \neq 0$)
$$\eqalign{
Q_{ii} =&{1\over (1-\omega)P_{ii}}\cr
0=&Q_{ij} (P_{ii} - \omega P_{jj})}\;\;\;.\eqno(2.12)$$
\noindent
We see  that the diagonal elements of $Q$ are related to those of $P$
and $Q_{ij}$ is arbitrary if $P_{ii}=\omega P_{jj}$,
otherwise zero. Finally we
can find as many inequivalent representations of the dimension as we want
for $H_\omega(3)$, and in general $P$ and $Q$ are not conjugated one
from the other.

\noindent
Of course, all that has been done here is equally valid for an arbitrary
$q$, but because, in this instance, $Q$ or $P$ are not necessarily
proportional to the
identity, other representations for $H_q(3)$ than those obtained
in eqs. (2.11) and (2.12) can be found. For $q=1$, however
it is known that $P$ or $Q$ can have a diagonal form, thus from (2.12) we see
that finite dimensional representations cannot be obtained. Finally, it should
mentionned that any invertible matrix $P$ leads to an appropriate
$Q=1/(1-q) P^{-1}$, but such a representation is not irreducible (see (2.10)).
\vskip 0.5truecm
d-- \underbar {The case of more than two generators}
\vskip 0.5truecm
\noindent
Returning to $ H_\omega$, it can be observed that along the same lines,
representations of the Heisenberg algebra generated by $p\; \theta$
and $\partial$, $H_\omega(2p+1)$, induced by generalized Grassmann algebra
can now be obtained.
But the interesting feature already
obtained in [14] is the possibility to get a matrix representation
of the quantum hyperplan, the natural structure which emerges by covariance
principles [20]. The price to be paid is however to use the singular
representation with $\omega^2$ when n is even ($\partial$ is expressed with
$\omega^2$ c.f. eq.(2.5)).

$$ \eqalign{
 \theta_i&=1^{\otimes(i-1)}\otimes\theta  \otimes \sigma_3^{\otimes(p-i)}\cr
 \partial_i&=1^{\otimes(i-1)}\otimes\partial\otimes \sigma_3^{\otimes(p-i)}}
\eqno(2.13.a)$$
$i=1 \cdots p$,
\vskip 0.3truecm
\noindent
 generate the algebra

$$\eqalign{
\theta_i \theta_j    &= \omega    \theta_j \theta_i     \,\,\, i < j \cr
\partial_i \partial_j&= \omega^{-1}\partial_j \partial_i \,\,\, i < j \cr
 \partial_i \theta_j&= \omega \theta_j \partial_i  \;\;\;\; i \neq j \cr
\partial_i \theta_i&=1+ \omega^2 \theta_i \partial_i +(-1+\omega^2) \sum
\limits_{ k>i}
 \theta_k \partial_k} \eqno(2.13.b)$$

\noindent
We see that the disymmetry between the $\theta_i$'s comes from (2.13.a)
and from $\partial \theta -\theta \partial = \sigma_3^2$.
\noindent

\vskip 0.4truecm
\noindent
Let us make some final remarks on this section. From (1.7), with $l=0$, we
can build the $\theta_i$ of (2.13.a), but, in contrast to the usual Clifford
case , we cannot obtain the $\partial_i$ from the $\theta_i^{(l)} \; (l\neq 0)$
because of (2.4) and (1.12). So,  whereas for the case $n=2$,
we can  build from
 $\theta_i=\theta_i^{(0)}$ and  $\partial_i= \theta_i^+=\theta_i^{(1)}$  the
spinorial representation of $O(2p)$, which is nothing  more than
$H_{-1}(2p+1)$,
when $n > 2$ the quantum hyperplan cannot be obtained from (1.7).
\vskip 1truecm
\noindent
3 -- \underbar{Finite dimensional representations of $sl(2)_q$}
\vskip 1truecm
\noindent
In this section, we want to construct some finite dimensional representations
of $sl(2)_q$ from $H_q(3)$. It should be observed that in [22], a
classification
of the finite dimensional representations of quantum groups has been addressed
when $q$ is not a primitive $n-$root of unity.
Deformations of $sl(2)$ have been classified  in [2], and among all those
deformations, we consider that which is introduced in [23], corresponding to
the second Witten's quantum deformation [24]. This type of algebra is
generated by $J^+,J^-,J^0$ which fulfil the relations
 $$\eqalignno{
& J^- J^0- q J^0 J^- = J^-\cr
& J^0 J^+- q J^+ J^0 = J^+&(3.1)\cr
& J^- J^+ -q^2 J^+ J^- = (q+1)J^0\cr}$$

\noindent
It is known that  $sl(2)_q$ is embedded into
  the $q-$deformed Heisenberg algebra $H_q(3)$ through the
following identification [25]

$$\eqalignno{
& J^+=(1-2\alpha(1-q))^{-1/2}(QQP -2\alpha Q)\cr
& J^0=(1-2\alpha(1-q))^{-1}\,\left\lbrack\left(1+{2\alpha(q^2-q)\over1+q}
\right)QP - {2\alpha\over1+q}
\right\rbrack&(3.2)\cr
&J^-=(1-2\alpha(1-q))^{-1/2} P\cr}$$
\noindent
where $\alpha$ is an arbitrary number and $P,Q$ are the generators of the
$q-$deformed Heisenberg algebra. So, particularly when $P,Q$ are
the $d-$dimensional
generators constructed in the previous part, a $d$-dimensional representation
for $sl(2)_q$ is obtained:
 any finite dimensional representations of $H_q(3)$ induce a
finite dimensional representation for  $sl(2)_q$. Of course, the converse
 is not true,
and infinite dimensional representation of  $H_q(3)$ may lead to a finite one
for $sl(2)_q$. That property is already true for the usual $sl(2)$
$(q=1)$, where
$d+1$ dimensional representation, corresponding to  polynomials of degree $d$,
could be obtained from  $H(3)$ through (3.2), in spite of the fact that the
latter
algebra does not admit finite dimensional representations [25].
This means that taking infinite dimensional matrix representations for $H(3)$,
using (3.2) we arrive at infinite matrices for the generators of  $sl(2)$, into
a block diagonal form, where blocks describe a finite dimensional
representation
for $sl(2)$.

\noindent
Returning to the construction of representations for $sl(2)_q$ induced by
$H_q(3)$,
we see that we get two kinds of representations:

(a) those coming from (2.11) and (2.12) for any $q$;

(b) those built from generalized Grassmann algebra when $q=\omega$.

\noindent
The first one is peculiar, as, using (2.11), we see that $J^-$ cannot be
 understood
as a usual annihilation operator because it is diagonal;
thus it is not nilpotent and
we do not have a highest weight state annihilated by $J^-$. Of course, such a
property is also valid for $J^+$. In fact the states which describe
the representation,
due to the special form of $J^-$, which is diagonal, are eigenvectors
of that annihilation operator:
thus in using  its action on any vector states we never span the whole
representation.

\noindent
This never occurs in the second type of representation,
because the generators
of generalized Grassmann algebras are nilpotent and so  $J^+,J^-$ .
They are thus
interpreted as usual creation and annihilation operators.

So using (3.2), we get various inequivalent representations of $sl(2)_q$
of the same dimension,
specified by the representation we choose for $H_q(3)$ and by $\alpha$.
Notice that when $2\alpha=\{d\}$ one has

$$(J^+)^{d+1} = (1-\{d\}(1-q))^{-(d+1)/2} q^{d(d+1)} Q^{2d+2} P^{d+1}
\eqno(3.3)$$

This property has already been exploited in ref.[25] to build $d+1$
dimensional representation
of $sl(2)_q$ from an infinite dimensional representation of $H_q(3)$,
in a similar way  as when one constructs
finite dimensional representation for  $sl(2)$. This representation is nothing
but the polynomial of degree $d$, where $Q=x$, the usual number, and
$P=D$, the Jackson
symbol for the derivative

$$Df(x) = {f(qx)-f(x)\over(q-1)x} \eqno(3.4)$$
It can be easily shown that $D$ fulfils the $q-$ graded Leibnitz rule.

Finally, it should be stressed again that unlike the usual Lie algebras, the
situation is quite different
for the $q-$ deformed algebras, because in the usual Lie algebras,
representations
of the type (a) does not exist. This property is essentially due to
the $q-$deformed
commutation relations, which allow 1-dimensional representations which
are not the trivial
ones.

\vskip 1truecm
4 -- \underbar{Conclusion}
\vskip 1truecm

We have studied all the finite dimensional representations of the
$q-$defor\-med Hei\-senberg
algebras, with  special attention to $q$ as an n-root of unity. The most
interesting representations are basically related to $n-$degree
polynomials, similar in the
way that fermionic oscillators are related to Grassmann variables.
This type of study should be useful, as $q-$deformed Heisenberg algebras
 provide
an appropriate tool for the construction of quantum algebras.
Furthemore, we can easily give an oscillator-type interpretation for
the $q-$ deformed Heisenberg algebras, and it can
be seen that $q-$deformations ($q-$oscillators) can be used in the framework of
quantum algebras in a way analogous to  the classical
harmonic oscillator. This feature has already been exploited in several
papers, with various extensions of the classical Heisenberg algebra [26]
(this algebra, after some transformations on the generators, is related to the
one we have studied); with
twisted second quantization for boson [20] ( which is
equivalent to the quantum hyperplan) or fermions [27]; by considering
the $q-$analogues of the Clifford and Weyl algebras [28] (an extension which
cannot be connected to generalised Clifford algebras or the quantum
hyperplan) and with superoscillators [29].
\noindent

\vskip 1truecm
Acknowledgment.
\vskip 1truecm
I would like to acknowledge N. Fleury  for helpful discussions, M. Rosso
for useful
remarks and advice especially in sect.2, and A. Turbiner who motivated this
work and was a great help for the work in sect.3.

\vskip 1truecm
\noindent
\centerline{REFERENCES}
\bigskip
\noindent
[1] V. G. Drinferl'd Proc. Int. Cong. of Mathematicians (MSRI, Berkley 1986)
798.
\hfill \break \null
[2] C.Zachos  {\it Paradigms of Quantum Algebras, Proceedings of the
 Conference
on Deformation
Theory of Algebras and Quantization with
 Applications to Physics}, Contemporary Mathematics, AMS
1991, J. Stasheff and M. Gerstenhaber (ed)
\hfill \break \null
[3] S. Majid Int. J. Mod. Phys. {\bf A5} (1990) 1.
\hfill \break \null
[4] C. N. Yang Phys. Rev. Lett {\bf 19} (1967) 1312; R. J. Baxter
 {\it Exactly Solved
Models in Statistical Mechanics}. Acad. Press, London, 1982; M. Jimbo
Int. J. Mod
Phys.{\bf A4} (1989) 3759.
\hfill \break \null
[5] L. Corwin, Y. Ne'eman and S. Sternberg Rev. Mod. Phys. {\bf 47} (1975) 573.
\hfill \break \null
[6] N. Roby C.R. Acad. Sc. Paris {\bf 268}(1969) 484
\hfill \break \null
[7] N. Roby Bull. Sc. Math. {\bf 94}(1970) 49
\hfill \break \null
[8] L. N. Childs Lin. and Mult. Alg.{\bf 5}(1978) 267
\hfill \break \null
[9] M. van den Bergh J. Alg. {\bf 109} (1987) 172
\hfill \break \null
[10] N. Fleury and M. Rausch de Traubenberg J.Math. Phys.{\bf 33}
(1992) 3356; {\it Finite dimensional representations of Clifford Algebras
of Polynomials} pre\-print CRN 94-03
  \hfill \break \null
[11] Ph. Revoy J. Alg. {\bf 46}(1977) 268
\hfill \break \null
[12] A. K. Kwasnievski J. Math.Phys. {\bf 26}(1985) 2284
\hfill \break \null
[13] D. B. Fairlie and C. K. Zachos Phys. Lett. {\bf B256} (1991)43
\hfill \break \null
[14] A.T. Filipov, A. P. Isaev and A. B. Kurdikov Mod. Phys. Lett.
{\bf A7} (1992) 2129
\hfill \break \null
[15] Y. Ohnuki and S. Kamefuchi,{\it Quantum Field Theory and Parastatistics
}, Univ.  of Tokyo press, 1982
\hfill \break \null
[16] V. Rubakov and V. P. Spiridonov Mod. Phys. Lett.{\bf A3} (1988)
1337; S. Durand, R. Floreanini and L. Vinet Phys. Lett {\bf 233B}
(1989) 523; S. Durand and L. Vinet
J. Phys. {\bf A23}  3661 (1990); N. Fleury , M. Rausch de Traubenberg
and R. M.
Yamaleev {\it Matricial Representation of Rational Power of Operators
and Paragrassmann Extension of Quantum Mechanics }, to appear in Int. J. Mod.
Phys. A.
\hfill \break \null
[17] A. B. Zamoloddchikov and V. I. Fateev Sov. Phys. JETP
 {\bf 62}(1985) 215; V. I. Pasquier and H. Saleur Nucl. Phys.
{\bf B330} (1990) 523.
\hfill \break \null
[18] K. Morinaga and T. Nono J. of Sc. Hiroshima Univ. Ser.
 {\bf A16}(1952) 13; A. O. Morris Quart. J. Math. Oxford
{\bf 18} (1967) 7; {\bf 19} (1968) 289.
\hfill \break \null
[19] H. Weyl {\it  The theory of Groups and Quantum Mechanics},
E. P. Dutton pp. 272-280, 1932 ( Reprinted, Dover, New York, 1950)
\hfill \break \null
[20] W. Pusz and S. L. Woronowicz Rep. Math. Phys {\bf 27} (1989) 231;
J. Wess and B. Zumino Nucl. Phys. (proc. Suppl.)
 {\bf 18B}(1990) 302
\hfill \break \null
[21] R. Balain and C. Itzykson C. R. Acad. Sc. Paris {\bf 303}(1986)
773
\hfill \break \null
[22] M. Rosso Com. Math. Phys. {\bf117} (1988) 581
\hfill \break \null
[23] O. Ogievetsky and A. Turbiner  $sl(2,R)_q$ {\it and Quasi Exact-Solvable
Problems}, Preprint CERN-TH 6212/91 ,1991 ( unpublished)
\hfill \break \null
[24] E. Witten Nuc. Phys. {\bf B330} (1990) 285.
\hfill \break \null
[25] A. Turbiner {\it Lie Algebras and Linear Operators with Invariant
Subspace},
To appear in {\it Lie Algebras  Cohomologies and New Findings in Quantum
Mechanics}
Contemporary Mathe\-matics, AMS,1993, N. Kamran and P. Olver (ed);
 A.Turbiner and
G. Post J. Phys. {\bf A27} (1994) L9;
N. Fleury and A. Turbiner {\it About the problem of Normal Ordering}
preprint CRN 94-08
\hfill \break \null
[26] A. J. Macfarlane J. Phys. {\bf A22} (1989) 4581;
L. C. Biedenharn J. Phys. {\bf A22} (1989) L873;P. P. Kulish and
E. V. Damaskinsky
J. Phys. {\bf A23} (1990) L415
\hfill \break \null
[27] W. Pusz Rep. Math. Phys {\bf 27} (1989) 349
\hfill \break \null
[28] T. Hayashi Com. Math. Phys. {\bf 127} (1990) 129
\hfill \break \null
[29]  M. Chaichain, P. Kulish and J. Lukierski Phys. Lett {\bf B262} (1991) 43
\bye